\documentclass[10pt,letterpaper]{article}
\usepackage{opex3}
\usepackage{siunitx}
\usepackage{multirow}
\usepackage{graphicx}
\usepackage{cite}
\usepackage{caption}

\begin{document}

\title{Characterization of the angular memory effect of scattered light in biological tissues}

\author{Sam Schott,$^{1,2}$ Jacopo Bertolotti,$^{1,3}$ Jean-Francois L\'{e}ger,$^{4,5,6}$ Laurent Bourdieu$^{4,5,6}$ and Sylvain Gigan$^{1,7,*}$}

\address{$^1$Institut Langevin, ESPCI ParisTech, CNRS UMR 7587, ESPCI, 1 rue Jussieu, 75005 Paris, France.\\
$^2$Cavendish Laboratory, University of Cambridge, J J Thomson Avenue, Cambridge CB3 0HE, UK.\\
$^3$Physics and Astronomy Department, University of Exeter, Stocker Road, Exeter EX4 4QL, UK.\\
$^4$Ecole Normale Sup\'{e}rieure, Institut de Biologie de l'ENS, IBENS, 46 rue d'Ulm, Paris, F-75005 France.\\
$^5$Inserm, U1024, Paris, F-75005 France.\\
$^6$CNRS, UMR 8197, Paris, F-75005 France.\\
$^7$Laboratoire Kastler Brossel, Universit\'{e} Pierre et Marie Curie, Ecole Normale Sup\'{e}rieure, CNRS, Coll\`{e}ge de France, 24 rue Lhomond, 75005 Paris, France.}

\email{$^*$sylvain.gigan@lkb.ens.fr} 


\begin{abstract}
High resolution optical microscopy is essential in neuroscience but suffers from scattering in biological tissues. It therefore grants access to superficial layers only. Recently developed techniques use scattered photons for imaging by exploiting angular correlations in transmitted light and could potentially increase imaging depths. But those correlations (`angular memory effect') are of very short range and, in theory, only present behind and not inside scattering media. From measurements on neural tissues and complementary simulations, we find that strong forward scattering in biological tissues can enhance the memory effect range (and thus the possible field-of-view) by more than an order of magnitude compared to isotropic scattering for $\sim$1\,mm thick tissue layers.
\end{abstract}


\section{Introduction}

Advances in deep tissue optical microscopy have been a driving force in biology and neuroscience for the last twenty years. Two-photon excited fluorescence microscopy has emerged as a powerful tool to study synaptic, cellular and integrative processes. However, scattering in neural tissues has limited optical imaging to superficial cortex layers only -- even after surgical removal of the skull \cite{Ntziachristos2010}. Deeper cortical layers (layers 4 to 6) are of substantial biological interest due to their interactions with other brain regions such as the thalamus \cite{Dombeck2010}. Together with subcortical structures, e.g, the hippocampus, they still remain inaccessible without removing the overlying tissue. Thus, even small improvements in imaging depth and resolution are bound to reveal novel insights on the functioning of mammalian brains. 

Efforts to increase the penetration depth of optical techniques have aimed to improve the quantity and focusing of ballistic photons by using regenerative amplifiers \cite{Beaurepaire2001}, longer wavelength lasers \cite{Horton2013} and adaptive optics \cite{Ji2012} as well as enhancing the collection of fluorescence photons \cite{Oheim2001}. Those efforts commonly sacrifice intensity in favor of resolution as the number of ballistic photons decreases exponentially with imaging depth. In an alternative approach, multiply scattered light itself can be used for imaging \cite{Vellekoop2008,Popoff2010}. Recently, the `optical memory effect' \cite{Feng1988} has emerged as a powerful tool to either enable the reconstruction of images from scattered light or to accelerate the imaging process\cite{Bertolotti2012,Katz2012,Nixon2013,Yang2014,Takasaki2014}. In essence, the memory effect (ME) describes an intrinsic isoplanatism of the scattering process for small angles even in strongly scattering media. By tilting the wavefront incident on a scattering medium, the emerging speckle patterns shifts accordingly and does not immediately decorrelate.

This isoplanatism can be exploited for adaptive optics imaging techniques. Spatial modulation of the incident wavefront permits focusing behind or inside turbid media if the transmission matrix is accessible (e.g., by placing a point source or detector behind the scattering layer) \cite{Nixon2013,Chaigne2013}. Since this comes at the cost of determining the transmission matrix for each image point, the ME plays a key role: if a focus is achieved, it can then be scanned over the ME range by simply introducing a linear phase tilt in the modulated wavefront \cite{Vellekoop2010}. 

In a more recent approach, a light-emitting object hidden behind an opaque layer is numerically reconstructed from its transmitted speckle pattern \cite{Katz2014}. Due to the ME, the autocorrelation of the speckle pattern is essentially equal to the object's autocorrelation and numerical inversion yields the object itself. This method is both non-invasive and does not require sophisticated imaging equipment but its field of view is intrinsically limited by the ME range.

Albeit important for imaging, the ME is in theory very limited and scales inversely with the thickness $L$ of the medium. Furthermore, in theory, it can not be observed \textit{inside} a scattering medium where the concept of isoplanatism breaks down, but only in transmission at a distance from the output plane. This can be understood when considering a tilt of the incident wavefront, i.e., the introduction of a linear phase gradient. The ME effect states that such a gradient, if sufficiently small, is preserved during the scattering process and results in the same linear phase gradient imposed on the emerging distorted wavefront. This gradient at the output plane then becomes a shift of the speckle pattern only after propagation through space. We thus expect to see the same decorrelation with the tilt of incident angle of the intensity pattern at the output plane itself but without the shift that is useful for imaging.

Recent work in biomedical imaging with adaptive optics \cite{Tang2012,Kong2014} suggests otherwise: phase corrections compensating for random scattering remain valid for a field-of-view of several microns inside brain tissues. This paper investigates how strong forward scattering inside biological media provides a much larger ME range that predicted by multiply scattering theory, in particular at intermediary depths of $\sim$1\,mm. Complementing angular correlation measurements in neural tissues, we simulate the impact of anisotropic scattering on wave propagation through multiple forward scattering layers and qualitatively reproduce the experimental results.

\section{Memory effect in anisotropic media}

In the limit of weak disorder $l\gg\lambda$ where $\lambda$ is the wavelength and $l$ the mean free path (MFP), the ME is well described by a first order approximation to ensemble averaged\footnote{Ensemble average denotes the average over possible realizations of disorder (here: locations of scatterers).} correlations in the intensity transmission matrix. In a medium much thicker than the MFP ($l\ll L$) the correlation $C$ obeys \cite{Feng1988,Berkovits1989}
\begin{equation}
C(\Delta\phi)=\left(\frac{k|\Delta\phi | L}{\sinh⁡(k|{\Delta\phi}| L)}\right)^2
\end{equation}
when following a tilt $\Delta\phi$ in both the incident wavefront and the emerging speckle pattern ($k=2\pi/\lambda$ denotes the wave number). As a figure of merit, we chose the angular difference $\Delta\phi_{1/5}$ at which the correlation drops to 1/5, the value at which visual resemblance to the reference image vanishes. This ME range scales linearly with the wavelength and is inversely proportional to the sample thickness: 

\begin{equation}
\Delta\phi^\textrm{theo}_{1/5}\approx 2.369\ k^{-1}L^{-1}.
\end{equation}

Note that Eq. (1) does not depend on the specific realization of the material such as the location of scatters or the transport mean free path so long as photons are scattered multiple times before reaching the output plane $l\ll L$. For a sample thickness  of $L = 1$\,mm and a wavelength of 532\,nm, the speckle correlation already drops to 1/5 at an angular shift of 11.5\,mdeg.

\begin{table}[t]
\begin{center}
\begin{tabular}{lcccl}
 Tissue& $\lambda$ [nm] & $l$ [\SI{}{\micro\meter}] &$l^*$ [\SI{}{\micro\meter}] & Reference\\ \hline
Slice&532&(26)$^\textrm{a}$&214&\cite{Mesradi2013}\\
Slice&775&55.2&(552)$^\textrm{a}$&\cite{Kobat2009}\\
Slice&800&(34)$^\textrm{b}$&(338)$^\textrm{b}$&\cite{Mesradi2013}\\
Slice, old rat &800&47&(470)$^\textrm{a}$&\cite{Beaurepaire2001}\\
Slice, juvenile &800&89&(890)$^\textrm{a}$&\cite{Beaurepaire2001}\\
Slice&1280&106.4&(1064)$^\textrm{a}$&\cite{Kobat2009}\\ \hline
In-vivo&775&131&(1310)$^\textrm{a}$&\cite{Kobat2009}\\
In-vivo&800&200&(2000)$^\textrm{a}$&\cite{Beaurepaire2001}\\
In-vivo&830&200&(2000)$^\textrm{a}$&\cite{Kleinfeld1998}\\
In-vivo&920&129&(1290)$^\textrm{a}$&\cite{Tang2012}\\
In-vivo&1280&285&(2850)$^\textrm{a}$&\cite{Kobat2009}, \cite{Kobat2011}\\
In-vivo&1700&365&(3650)$^\textrm{a,c}$&\cite{Horton2013}\\ \hline
\end{tabular}
\end{center}
\caption{Scattering properties of in-vivo and ex-vivo cortex tissues at different wavelengths. Values are obtained from references in the last column and values in parenthesis are deduced from the experimental data with (a) $g=0.88$ at 532\,nm and $g=0.9$ for $\lambda>775$\,nm \cite{Yaroslavsky2002}, (b) by extrapolating the fit in \cite{Mesradi2013} up to 800\,nm and (c) by neglecting water absorption.}\label{tab1}
\end{table}

The situation is expected to improve somewhat for biological tissues where the above conditions are no longer valid since waves are scattered preferentially in the forward direction in biological samples. This anisotropy is characterized by the average scattering angle
\begin{equation}
g=\int_{4\pi}p(\theta)\cos\theta\textrm d \Omega\ .
\end{equation}
Here, $p(\theta)$ denotes the probability of a photon being scattered into the angle $\theta$ relative to its incident direction with $\int_{4\pi} p(\theta)\textrm d \Omega=1$ and the solid angle $\Omega$. It is easy to see that isotropic scattering yields $g=0$ while complete forward scattering, i.e., no scattering at all, gives $g=1$. In the presence of such anisotropy, the relevant transport parameter is no longer $l$ but the transport mean free path (TMFP) $l^*$ with $l^* = l/(1-g)$ \cite{Akkermans2007}. While the $l$ is the average distance between scattering events, $l^*$ can be interpreted as the distance at which the direction of wave propagation has become independent from the initial direction. The condition for Eq. (1) now reads $l^*\ll L$ and is easily broken for large $g$-factors and intermediate sample thicknesses. This applies in particular to biological tissues that typically show large g-factors between 0.8 and 0.98 \cite{Cheong1990}. Scattering parameters of the rodent cortex from literature are given in Table \ref{tab1} for in-vivo and ex-vivo measurements. In that regime, the ME range is presumed to be larger than predicted by Eq. (1) and possibly present inside scattering samples as well where some directionality is preserved for distances inferior to $ l^*$. 
However, the ME in this forward scattering regime is not well understood and we are not aware of any experimental studies to date linking speckle correlations with scattering anisotropy.

Here, we present a systematic study of the ME in chicken breast and rat cortex samples. We identify the anisotropy of scattering as a key factor for the extended correlation range by matching experimental results with a simulation of wave propagation through multiple forward scattering layers. In order to get a qualitative understanding of this effect, we introduce an effective thickness $L_\textrm{eff}$ that corresponds to the equivalent thickness of an diffusive medium producing the same memory effect. In essence, a larger memory effect corresponds to a thinner effective sample thickness.

\section{Experiment}

\begin{figure}[b]
 \centering
 \includegraphics[width=9cm]{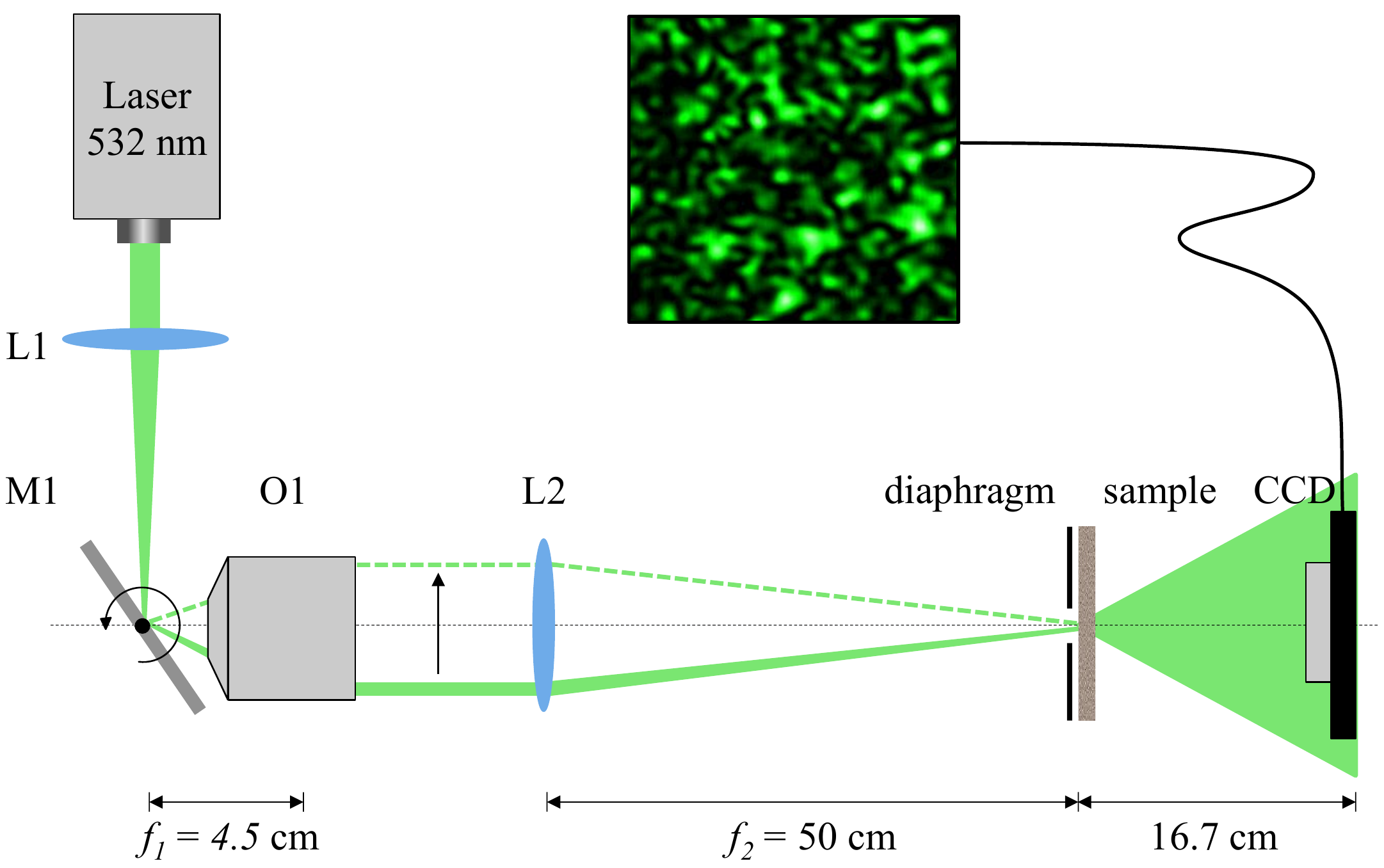} 
 \caption{Experimental setup to measure the optical memory effect: A 532\,nm laser beam is focused on the axis of a mirror (M1) mounted on a rotation stage. This spot is imaged onto the sample through an objective (O1) with focal length $f_1=4.5$\,cm and a lens (L2) with focal length $f_2 = 50$\,cm. This results in a circular sample illumination area with 2.3\,mm diameter. The scattered light is captured by a CCD with $1024\times1280$ pixels where one pixel measures $5.3\times\SI{5.3}{\micro\meter}^2$. An optional diaphragm can be inserted in front of the sample to reduce the beam size.}
 \label{2}
\end{figure}

\begin{figure}[b]
 \centering
  \includegraphics[width=10cm]{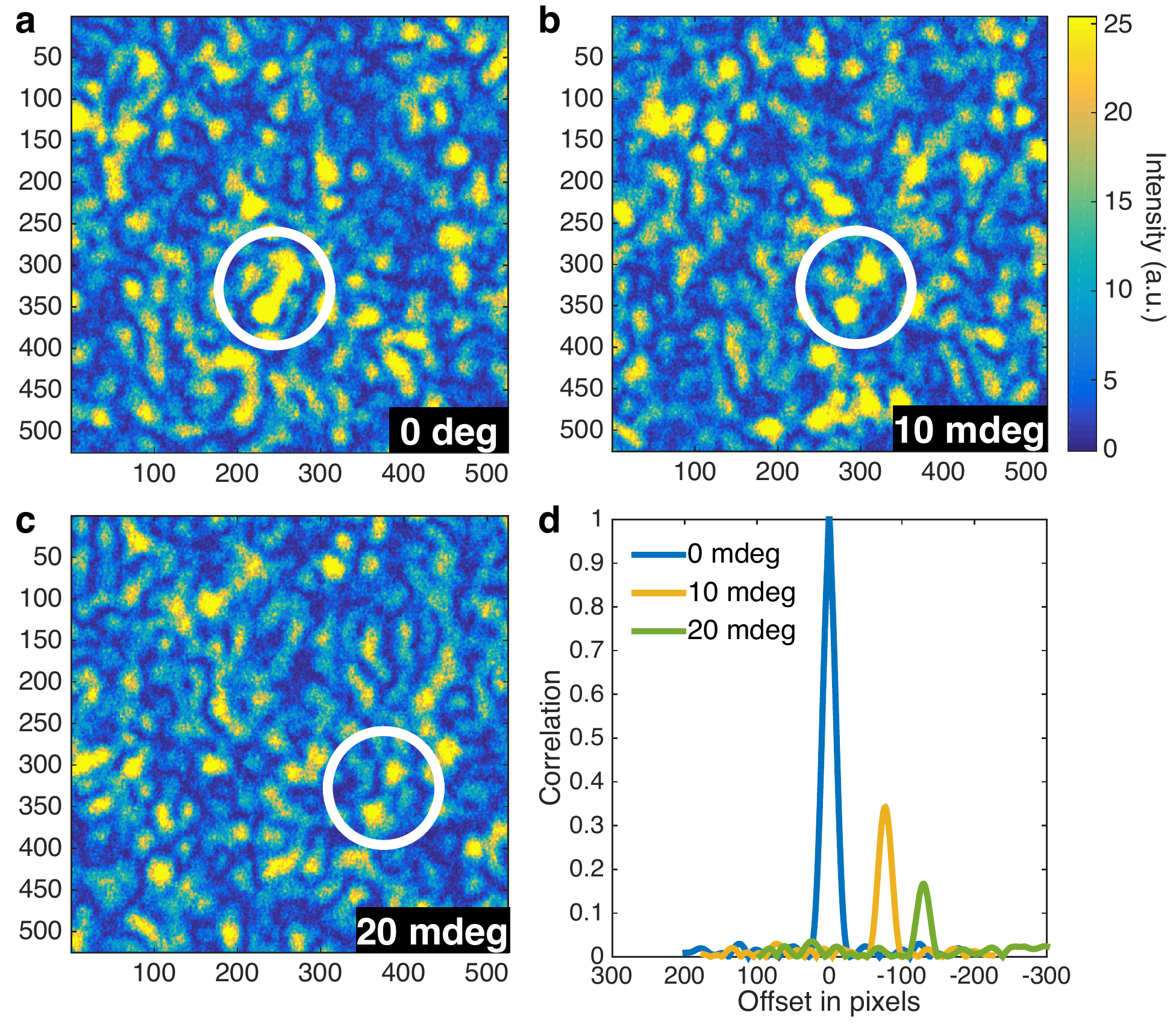} 
  \caption{(a) to (c) show speckle patterns excerpts from rat cortex samples at three different angles. Axes denote pixels from CCD camera, intensity in arbitrary units. Horizontal shift and decorrelation are visible when following the circled speckle. (d) Calculated correlations against offset for images (a) to (c) with 0\,deg as the reference image.}
  \label{1}
\end{figure}

A sketch of the experimental setup for ME measurements is shown in Fig. \ref{2}. To measure the decorrelation of speckle patterns with changing incident angle, we focus a continuous wave laser beam (wavelength, 532\,nm) on the rotation axis of a mirror mounted on a motorized rotation stage and image that spot onto the scattering sample. The lenses’ focal lengths were chosen to provide 100/9 fold magnification and a corresponding increase in angular resolution. A CCD camera placed in the far field behind the sample then captures the transmitted light. By taking a series of images while rotating the mirror at constant speed, we capture speckle patterns for a series of incident angles with a fixed point of illumination. Such speckle patterns are shown for three incident angles in the case of brain slices in Fig. \ref{1}a-c. The correlation between two images is then calculated as a function of possible offsets between patterns, giving the actual correlation to its reference from the maximum value (Fig. \ref{1}d).

Measurements on a ground-glass diffuser and zinc oxide (ZnO) scattering sample were performed to confirm alignment and accuracy of the experimental setup. Albeit being completely opaque, ground glass diffusers effectively consist of a single scattering layer and thus have a very large ME range. In contrast, ZnO is an ideal multiply scattering sample: it does not absorb visible light but scatters it quite effectively with a refractive index contrast, $\Delta n\approx 2$ \cite{Coutts2006}. Correlation curves of a ZnO diffuser are well reproduced by multiple scattering theory (Fig. 3a) and the fit parameter $L_\textrm{eff}=\SI{648}{\micro\meter}$ is close to the actual layer thickness of $L=\SI{650}{\micro\meter}$. This gives us the expected small ME range of $\Delta\phi_{1/5}=17.5$\,mdeg.

Additionally, we confirmed that the angular correlation curves are independent of illumination spot size and thus speckle size for both isotropic and forward scattering (Fig. S1).

As model biological systems, we chose chicken breast muscle tissue and rat cortex slices, the first because it is readily available and the second since it is an actively researched system in neuroscience. To obtain an imaging geometry similar to in vivo experiments, three month old Wistar were transcardially perfused with 4\% paraformaldeide (PFA) in phosphate buffer. The cortices of extracted brains were then flattened and stored overnight in a solution of 4\% PFA between two glass slides separated by 1.5\,mm thick spacers. Tangential slices of different thicknesses were cut from the flattened cortices.
We mounted rat cortex samples in water and performed all measurements within 1-2 days after preparation while chicken breast samples were mounted as is. Respective correlation curves are shown in Fig. 3 and both effective thickness and ME range values are given in Table \ref{tab2}.

The most striking observation is that the fit parameter $L_\textrm{eff}$ is around one order of magnitude smaller than the actual thickness $L$, the ME range is thus much larger than for a multiple scattering sample of the same thickness.

\begin{table}[b]
\begin{center}
\begin{tabular}{p{1mm}p{1mm}rrrr}
 &&$L$ [\SI{}{\micro\meter}] &$L_\textrm{eff}$ [\SI{}{\micro\meter}] & $\Delta\phi_{1/5}^\textrm{theo}$ [mdeg] & $\Delta\phi_{1/5}^\textrm{exp}$ [mdeg] \\ \hline
\multicolumn{2}{c}{ZnO}& 650 & 648 & 17.7 &17.5 \\ \hline
\multirow{5}{*}{\rotatebox{90}{Chicken}}&\multirow{5}{*}{\rotatebox{90}{breast}}
& 850 & 29$\pm$1 & 13.5 & 476 \\
& & 1000	 & 31$\pm$1 & 11.5 & 438 \\
& & 1170 & 69$\pm$3 & 9.8 & 203\\
& & 1660 & 93$\pm$9 & 6.9 & 208 \\
& & 2730 & 570$\pm$10 & 4.2& 21\\\hline
\multirow{5}{*}{\rotatebox{90}{Rat cortex}}&\multirow{3}{*}{\rotatebox{90}{top}}& 400 & 45$\pm$9 & 28.7 & 301\\ 
&& 800 & 388$\pm$8 & 14.3 & 30\\
&& 1600 & 650$\pm$20 & 7.2 & 17\\ \cline{2-6}
&\multirow{2}{*}{\rotatebox{90}{mid}}& 400 & 148$\pm$30 & 28.7 & 98 \\ 
&& 800 & 335$\pm$8 & 14.3&34 \\ \hline
\end{tabular}
\end{center}
\caption{Effective thickness $L$ and ME range for ZnO, chicken breast and rat cortex tissue. Uncertainties in sample thickness $L$ are about \SI{50}{\micro\meter} for tissues and \SI{8}{\micro\meter} for ZnO while error ranges for $L_\textrm{eff}$ are fitting uncertainties.}\label{tab2}
\end{table}

For chicken muscle tissue, a TMFP of $l^* = \SI{1.25}{\milli\metre}$ (MFP, $l = \SI{43.7}{\micro\metre}$) and an anisotropy factor of $g = 0.965$ have been measured \cite{Cheong1990}. Our sample thicknesses were chosen to range from less than one $l^*$ to above $2l^*$. But due to the inherent inhomogeneity of biological tissues, those values are likely to fluctuate both within and between samples. This becomes evident when comparing the \SI{1170}{\micro\metre} and \SI{1660}{\micro\metre} slices: although the thickness increases, we observe that the ME ranges slightly decreases. Keeping this variability in mind, there is still a clear trend emerging from the chicken breast measurements. The ME range is up to 35 times larger than expected for isotropic scattering but this difference decreases to a factor of 5 for the thickest sample. Eventually, it is expected to vanish entirely for samples much thicker than $l^*$, i.e., in the cm range. In addition to the extended ME range, we also observe a deviation from the theoretical $(x/\sinh x)^2$ bell shape of angular correlation curves towards a more exponential shape. For the thickest sample however, the ideal theory shape is recovered.

\begin{figure}[b]
  \centering
  \includegraphics[width=10cm]{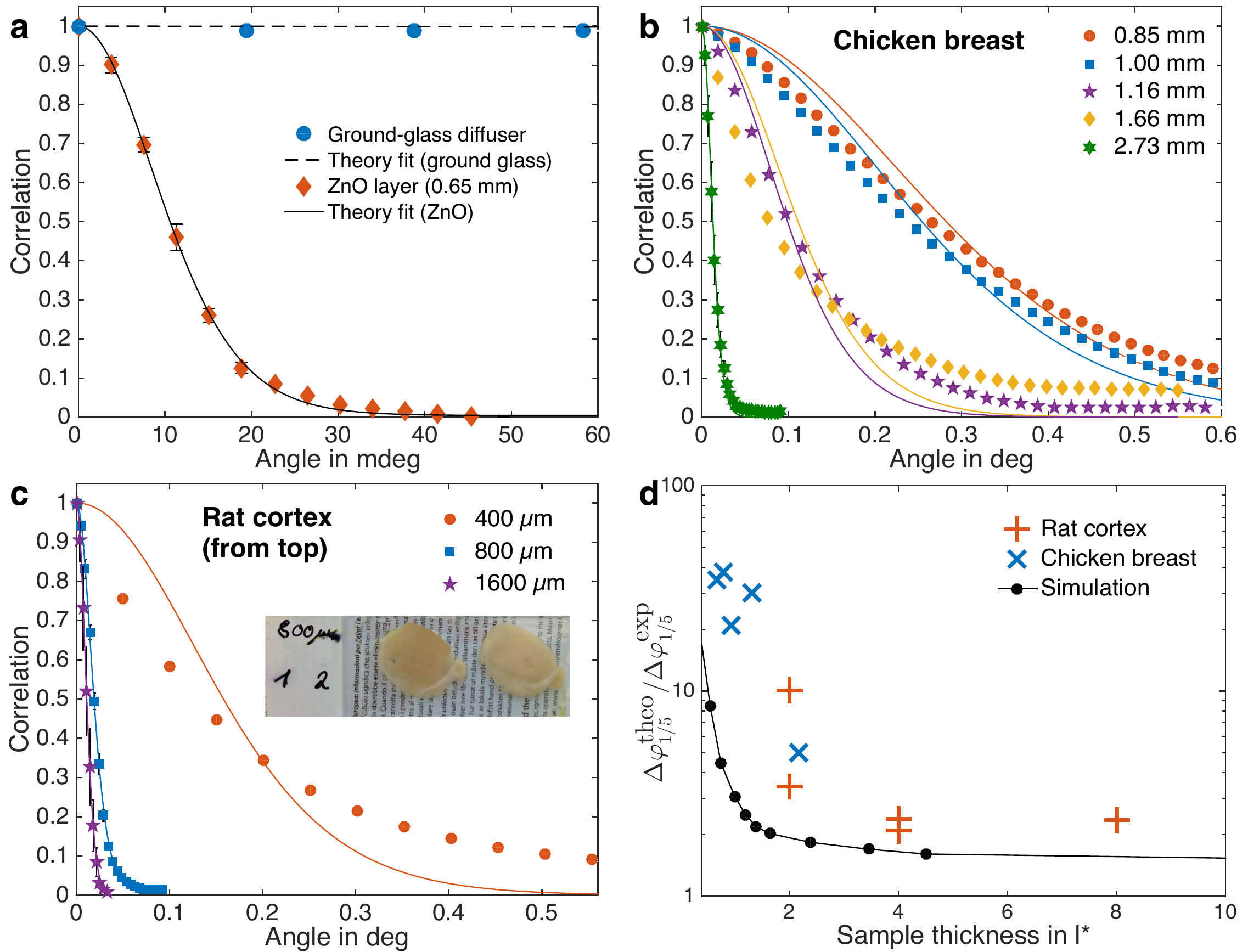} 
  \caption{Angular correlation functions for (a) 0.65\,mm thick zinc oxide layer and ground glass diffuser as references, (b) chicken breast slices with thicknesses from 0.85\,mm to 2.73\,mm and (c) rat cortex slices from first series. Solid lines are fits with Eq. (1) and error bars (standard deviations over multiple measurements) are omitted if smaller than the marker size. (d) Ratio between experimental and theoretical (isotropic scattering) ME range for different sample thicknesses in multiples of $l^*$. Thickness values have an uncertainty of $\sim20$\% from variations in the TMFP. Simulated values are shown as black circles with simulated ratio of 1.08 at $50l^*$ (outside the plotted range).
  }
  \label{3}
\end{figure}

A similar picture emerges from the rat cortex samples. Two series of cortex slices were prepared to account for brain layers with different properties. The first series starts at the cortex surface (layer 1) and is labeled `top' in Table \ref{tab2} while the second starts at the cortex middle (label `mid') and the thickest sample consists of an entire flattened cortex. With a smaller TMFP of $200\pm50$\,\SI{}{\micro\metre} \cite{Nishidate2011,Mesradi2013} in rat cortices, all of our samples are thicker than $l^*$. Nevertheless, we observe the same qualitative differences in ME shape and range as for chicken tissue. We recover the ideal theory shape for angular correlation curves for the two thicker samples and at a thickness of $d\approx8l^*$, the experimental ME range is only twice as big as the theory value. 

The change in ME curve shape indicates that the additional correlations are of a different nature than in the conventional ME. In multiply scattering media, the ME emerges from correlations between the input and output positions for light: a pencil beam incident on a thin slab spreads through the medium and emerges as a diffuse spot at the output plane, or a size roughly given by the thickness, but with a total loss of the incident direction. It corresponds to a macroscopic structure in the transmission matrix, with larger transmission amplitudes close to the diagonal in position space\cite{Judkewitz2014}. But in case of strong forward scattering and a sample thickness close to the TMFP, the direction of wave propagation is not completely randomized when passing through the sample. Instead, as Judkewitz et al. argue \cite{Judkewitz2014}, one input mode in k-space is transmitted only to a narrow cone of output directions around the incident k, thereby giving the transmission matrix a similar structure with large near-diagonal amplitudes in k-space. They derive translational (instead of angular) correlations arising from that k-space structure, even for a negligible traditional ME, i.e., for very thick slabs. In practice for our sample, we expect those correlations to contribute to the angular ME in forward scattering tissues as well.

In the above discussion, we have neglected the effects of absorption on the ME. In fact, absorption does increase angular correlations by narrowing the diffusive cone that would emerge from a pencil beam (larger near-diagonal amplitudes in the real space transmission matrix). Van Rossum and Nieuwenhuizen \cite{VanRossum1993} derived a modified version of Eq. (1) depending on the absorption coefficient $\kappa$ but resulting increases in correlation are very small even for strongly absorbing materials with $\kappa = L/2$ and retain the shape of Eq. (1). Therefore, we can safely exclude absorption effects as the origin of larger correlations.

Another question is whether correlations observed in the far field are retained close to the output plane where only a fraction of the scattered amplitudes contribute to the interference pattern. This is of special interest for imaging inside biological tissues where we  ideally would like to scan a focus or exploit the memory effect inside a medium. We mimicked a ‘near field’ situation by placing a diaphragm immediately behind the sample output plane. Resulting correlations in the far field speckles thus correspond to angular correlations at the aperture. The obtained correlation curves were indeed the same as without diaphragm (Fig. S2). However, in a strongly multiple scattering sample, we do not expect an actual shift of the output plane intensity pattern since the ME relies on well-defined output modes in k-space. Instead, we would expect a mere decorrelation that only becomes a shift in the far field. While this control experiment is not a definite confirmation that ME is present inside tissues, we believe the combination of lower effective thicknesses together with the presence of translational invariance as measured in \cite{Judkewitz2014} is in agreement with the experimental observation of the ME inside tissues exploited in \cite{Tang2012,Kong2014} for imaging.

\section{Phase mask simulation}

\begin{figure}[b]
  \centering
  \includegraphics[width=10cm]{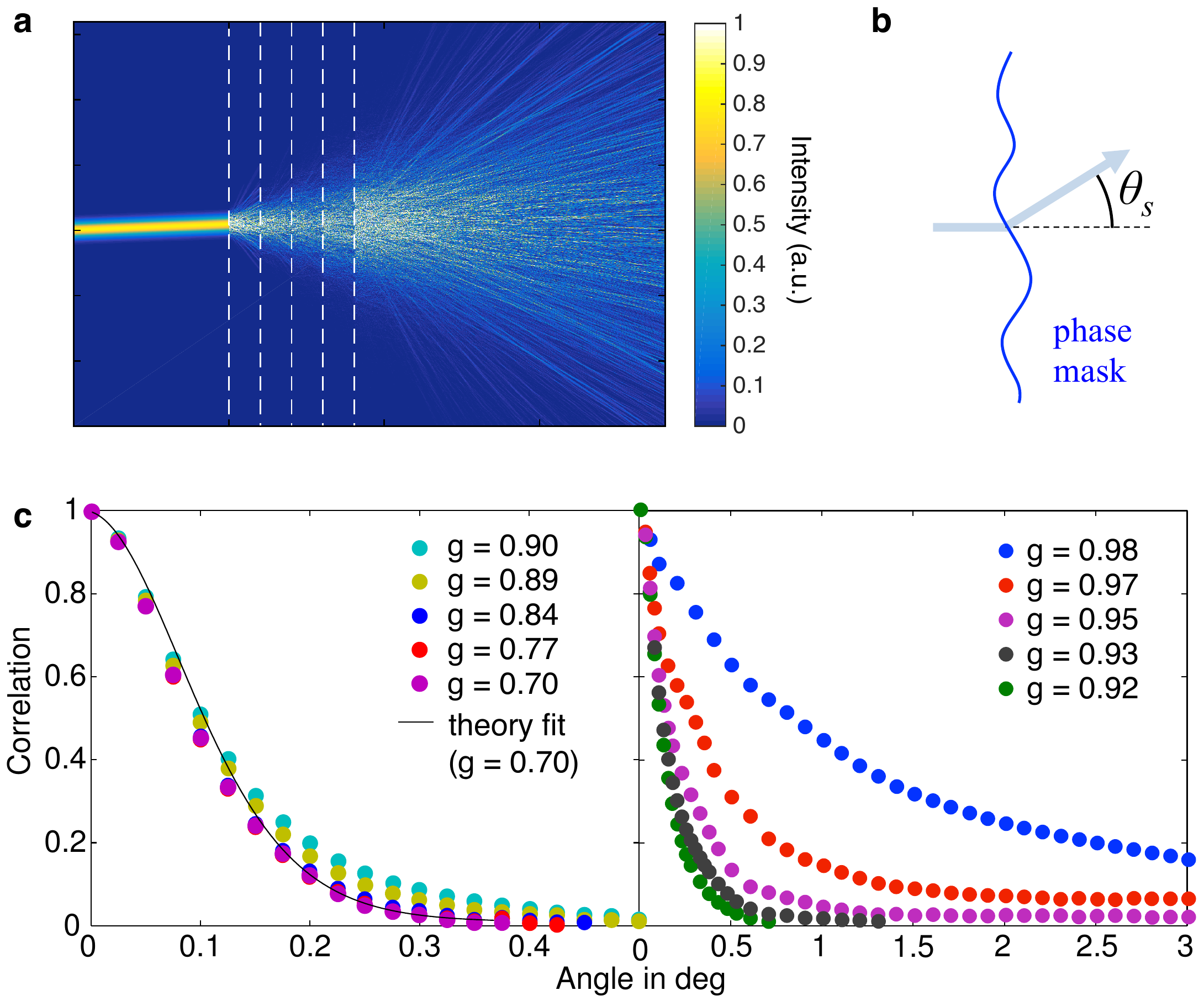} 
  \caption{(a) 2D cross-section of the light intensity in a simulation with five phase masks (dashed white lines), incident angle of 3\,deg. (b) Cross-section of the path difference $\lambda\Phi(x,y)/2\pi$ from a random phase mask. The resulting scattering angle $\theta_s$ is determined by the phase mask gradient. (c) Simulated angular correlation curves in the weak forward scattering regime. The sample consists of two phase masks with a distance of $100\lambda$ and $g$-factors from 0.7-0.98.}
  \label{4}
\end{figure}

To confirm the scattering anisotropy as cause for a longer ME range, we aim to reproduce the impact of larger g-values on correlation curves in a multiple scattering simulation. Again, we are especially interested in optically thin samples with predominant forward scattering and therefore model the scattering process with a number of consecutive phase masks (Fig. 4a). Each of those phase masks $\Phi(x,y)$ is created from a random matrix by multiplying its Fourier spectrum with a 2D Gaussian function and thus determining the frequencies of its spatial fluctuations. By changing the Gaussian's width, we change the scattering angles $\theta_s$ and thus the $g$-factor via (Fig. 4b)

\begin{equation}
\theta_s (x,y)=\arctan \left(\frac{\lambda}{2\pi} |\nabla \Phi(x,y)|\right)
\end{equation}

At a phase mask $n$, the scalar electric field $E$ accumulates a phase of $e^{i\Phi_n (x,y)}$ and we use the Rayleigh-Sommerfeld equation \cite{Marathay2004}, an extract solution to the scalar Helmholtz equation, for light propagation.
There are two obvious shortcomings of this simplified model: First, phase masks allow for forward scattering only and all back scattering is thus assumed to be negligible and second, scattering events take place only at certain planes unlike the reality of randomly distributed scatters with anisotropies on a large range of length scales \cite{Mourant2000}. We trade accuracy for shorter computation times but nevertheless are able to control MFP (distance between phase masks), $g$-factor and TMFP of the scattering process.

Keeping the above in mind, we still see good qualitative agreement in both ME range and curve shape. We simulated correlation curves for anisotropy factors from 0.70-0.98 while keeping all remaining parameters fixed. As expected, we see the strongest impact of g on correlation curves in case of only a few scattering events and strong forward scattering (Fig. 4 c,d). We qualitatively reproduce the curve shape for high $g$-values and retrieve the bell-like shape from multiple theory at lower $g$-values.
In a second series of simulations, we kept the MFP (distance between phase masks) fixed at \SI{43}{\micro\metre} while varying the total sample thickness from ~$0.1l^*$ to $50l^*$. The resulting ME ranges are shown in Fig. 3d together with experimental values. As expected, the ratio between simulated forward scattering and theoretical multiple scattering ME range again increases with thinner samples. However, the simulation somewhat underestimates the impact of anisotropy and $\Delta\phi_{1/5}^\textrm{exp}/\Delta\phi_{1/5}^\textrm{theo}$ is uniformly bigger than the simulated values. This comes as a surprise since we expect the exclusion of backward scattering to produce stronger correlations. The discrepancy might originate from the discrete scattering planes in our model. Usually, distances between scattering events are distributed around the MFP and some photons undergo less scattering events than others (‘snake photons’). The phase mask simulation assumes the same number of scattering events for all photons and might thus reduce speckle correlations.

Even though angular correlations are underestimated, we can still qualitatively reproduce the cusped curve shape for weak forward scattering in thin slices together with the transition to the theory shape of Eq. (1) for stronger scattering in thicker samples.

\section{Conclusion}
Several emerging imaging techniques for scattering media rely on angular correlations of the transmitted light \cite{Bertolotti2012,Katz2012,Katz2014}. We have shown here that anisotropic scattering can extend the range of this optical memory effect by more than an order of magnitude when propagation through tissues. 

Both the experiments and our phase mask simulation show that the preservation of directionality during scattering results in correlations that become visible in the distinct exponential like shape of angular correlation curves. From our measurements on chicken and rat tissues, we expect that diffraction-limited imaging with a field-of-view of 50-100\,mdeg through a tissue layer of 1\,mm should be possible at wavelength of 533\,nm. The ME range scales linearly with $\lambda$ and the TMFP is known to increase by factor of 5-10 in the near infrared and for in-vivo compared to in-vitro slices (see Table \ref{tab1}). At common wavelengths used in two-photon fluorescence microscopy, the ME range should increase accordingly.

Further improvements might be achieved through several methods. First, translational correlations that arise from anisotropic scattering \cite{Judkewitz2014} could well be exploited together with temporal and spectral correlation alongside the traditional angular ME to further increase the field-of-view. Second, we have seen that the ME is strongest for photons that have undergone only a limited number of scattering events. Combining traditional gating techniques that only retain snake photons \cite{Search} with imaging processes that use scattered light should yield further improvements for both field-of-view and imaging depth.

With an emerging better understanding of speckle correlations in anisotropic media, we expect imaging techniques that utilize scattered photons to become far more powerful for applications in biological imaging or photostimulation.

\section*{Acknowledgements}
This work was funded by European Research Council Grant 278025 and the Agence Nationale de la Recherche (Investissements d'Avenir ANR-10-LABX-54 MEMO LIFE, ANR-11-IDEX-0001-02 PSL* Research University). We thank Prof. Georg Maret for enabling Sam Schott's stay at institut Langevin and his support of the project and David Martina for technical help in the development of the experimental setup.

\bibliographystyle{osajnl.bst}

\newpage

\section*{Supplementary figures}

\begin{figure}[h]
 \centering
  \includegraphics[width=10cm]{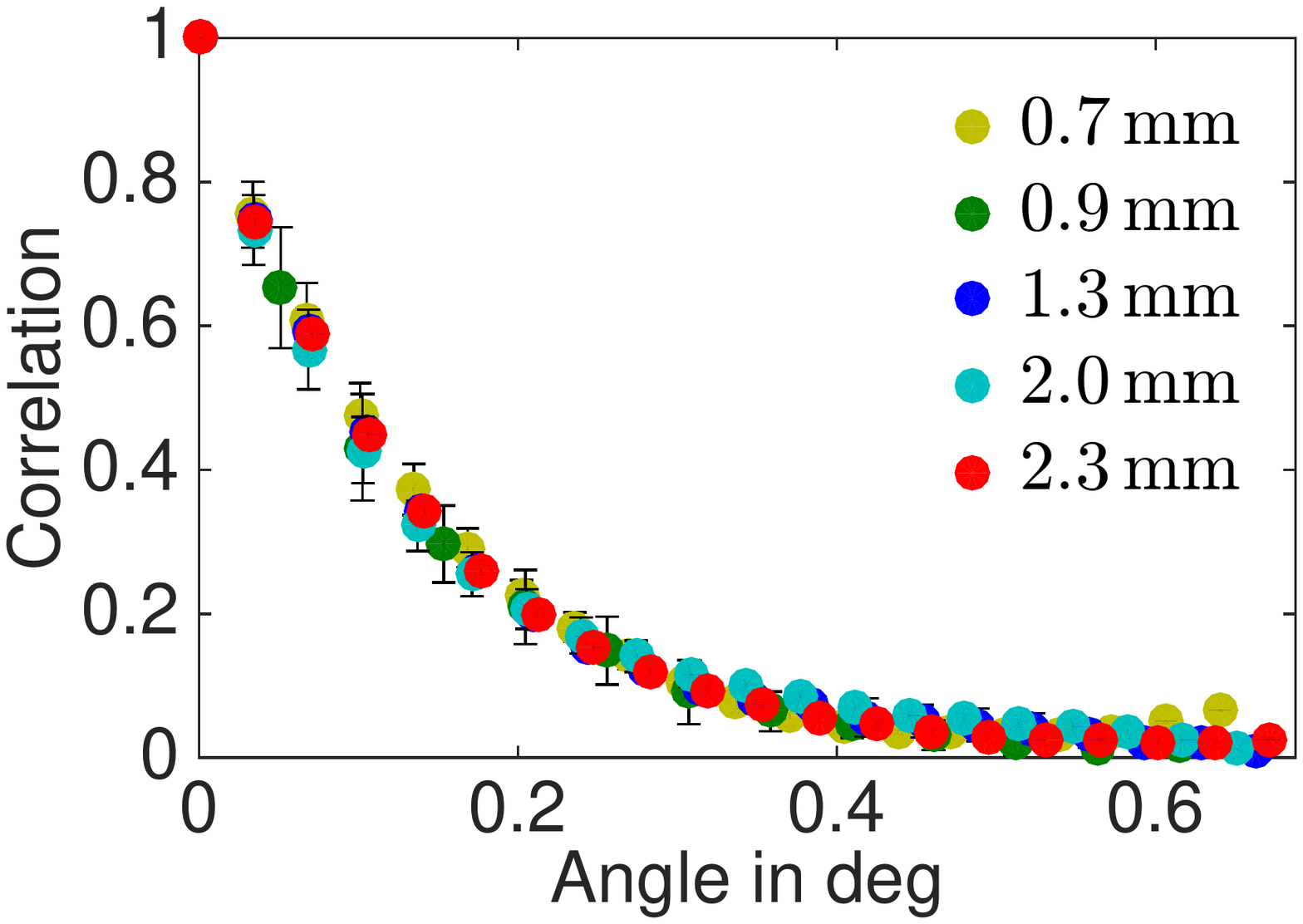} 
  \vspace*{-1.5in}
  \caption*{Fig. S1. Influence of beam size on angular correlation function and speckle pattern for a \SI{400}{\micro\meter} thick rat cortex sample. The angular correlation function remains the same for spot sizes from 0.7\,mm to 2.3\,mm while the average speckle size on the CCD camera (determined from the speckle pattern's autocorrelation) is reduced by half. The beam size is adjusted with an aperture in front of the sample.}
  
  \label{S1}
  \end{figure}
  
  \begin{figure}[t]
   \centering
   \vspace*{-1.4in}
  \includegraphics[width=10cm]{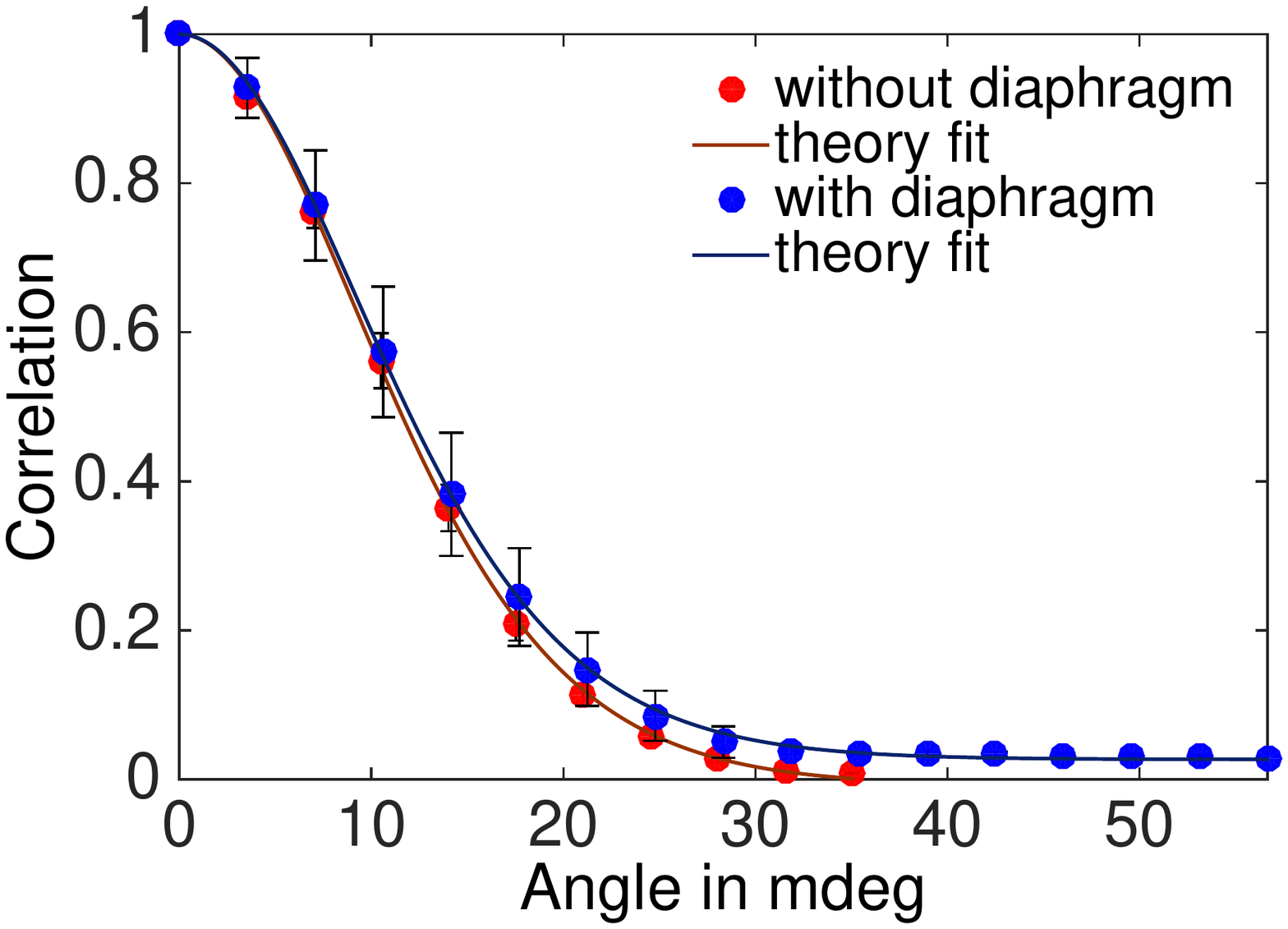} 
  \vspace*{-1.4in}
  \caption*{Fig. S2. Accessing the ``near field'' by placing an aperture (diameter, 0.7\,mm) immediately behind the sample exit plane. The speckle pattern is then formed by interfering light from a limited area only. This allows us to access the angular correlation of transmitted light in the aperture plane. While the angular correlation function remains the same, the speckle pattern changes and the speckle size increases from \SI{18}{\micro\meter} to \SI{65}{\micro\meter}. The emerging plateau is an artifact from the larger speckle size: random similarities are weighted more strongly with a smaller number of speckles per image. Measurements were performed on a \SI{1600}{\micro\meter} thick rat cortex sample.}
  \label{S2}
\end{figure}

\end{document}